\documentclass[apjl]{emulateapj}
\bibpunct{(}{)}{;}{a}{}{,}

\begin{document}
% Title Page
\title{Energy release from impacting prominence material following the 2011 June 7 eruption}

\author{H. R. Gilbert, A. R. Inglis\altaffilmark{1}, M. L. Mays\altaffilmark{1}, L. Ofman\altaffilmark{1}, B. J. Thompson, C. A. Young, }
\affil{Solar Physics Laboratory, Heliophysics Science Division, NASA Goddard Space Flight Center, Greenbelt, MD, 20771, USA}

\altaffiltext{1}{Physics Department, The Catholic University of America, Washington, DC, 20664, USA}

\begin{abstract}

Solar filaments exhibit a range of eruptive-like dynamic activity, ranging from the full or partial eruption of the filament mass and surrounding magnetic structure as a coronal mass ejection (CME), to a fully confined or ‘failed’ eruption. On 2011 June 7, a dramatic partial eruption of a filament was observed by multiple instruments on SDO and STEREO.  One of the interesting aspects of this event is the response of the solar atmosphere as non-escaping material falls inward under the influence of gravity. The impact sites show clear evidence of brightening in the observed EUV wavelengths due to energy release. Two plausible physical mechanisms explaining the brightening are considered: heating of the plasma due to the kinetic energy of impacting material compressing the plasma, or reconnection between the magnetic field of low-lying loops and the field carried by the impacting material. By analyzing the emission of the brightenings in several SDO/AIA wavelengths, and comparing the kinetic energy of the impacting material (7.6$\times10^{26}$ - 5.8 $\times10^{27}$ ergs) to the radiative energy ($\approx$ 1.9 $\times 10^{25}$ - 2.5 $\times 10^{26}$ ergs) we find the dominant mechanism of energy release involved in the observed brightening is plasma compression. 

\end{abstract}

\maketitle

\section{Introduction}
\label{intro}

Solar filaments (called prominences when observed on the solar limb) exhibit a range of eruptive behavior, including dramatic activation with the filament mass remaining confined to the low corona \citep[e.g.][]{2003ApJ...595L.135J, 2006ApJ...653..719A}, the eruption of part of the observed filament structure \citep{2006ApJ...651.1238Z}, and the almost complete eruption of all of the filament mass \citep{2000SoPh..194..371P}.  The most common type of eruption is the partial eruption, where prominence mass is observed falling back to the solar surface \citep[see][for a thorough discussion on different types of eruptions in the context of kinking motions]{2007SoPh..245..287G}.  Prominence eruptions (full and partial) are often associated with coronal mass ejections (CMEs) and may play an important role in their initiation.  

Partial eruptions are particularly interesting because of what happens after the eruptive part of the material escapes; the remaining supporting magnetic structure and prominence mass may relax or return to a lower altitude after the magnetic field reconfigures.  This leads to the question: what happens to the returning mass?  Observationally, partial eruptions are most obvious on the limb, where pieces of material can be seen falling along apparent magnetic field lines in the plane of the sky \citep{2001ApJ...549.1221G}, but are occasionally observed on the solar disk \citep{2002SoPh..207..111P, 2012ApJ...753...52L}. Another phenomenon associated with filament eruptions that has historically been linked to returning material is the two-ribbon flare.  One early explanation for two-ribbon flares is provided by \citet{1967SoPh....2...49H} in which the author concludes that two-ribbon flares are due to a chromospheric flare-like brightening mechanism. More recently, \citet{2002ApJ...567L..85S} proposes plasma falling from high altitudes after filament eruptions can convert potential energy into thermal energy, providing the source of energy and mass supply for long-duration solar flare events in the long-lasting decay phase.  

In the present work, we are interested in the physics behind the interaction of falling material and the solar atmosphere upon impact, which have the observational signature of brightening. Other phenomena with a similar observational signature are sequential chromospheric brightenings ( SCBs) \citep{2005ApJ...630.1160B, 2007AdSpR..39.1781P, 2012ApJ...750..145K}.  \citet{2012ApJ...750..145K} propose that SCBs are caused by the chromospheric impact of accelerated plasma along newly reconfigured magnetic field lines.  Further research will reveal whether the SCBs are related to the process described in this paper, or if they are a fully independent phenomenon.

\begin{figure*}
\begin{center}
\includegraphics[width=16cm]{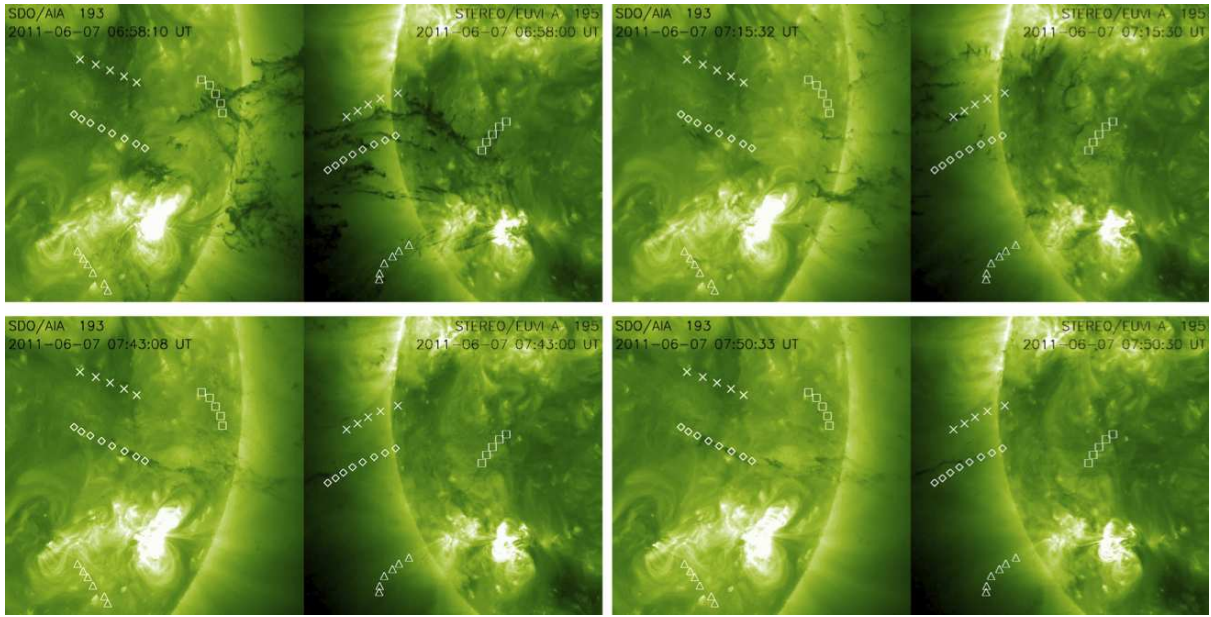}
\caption{SDO/AIA 193 $\AA$ and STEREO-A/EUVI 195 $\AA$ images with overlays showing the trajectories of four pieces of falling filament material following the 2011 June 7 flare. The context images are shown at four time intervals, close to the respective impact times of each piece. In each panel the track corresponding to impact region 1 is denoted by diamonds, region 2 by triangles, region 3 by squares and region 4 by crosses.  }
\label{fig1}
\end{center}
\end{figure*}

The break-up of returning plasma after the 2011 June 7 filament eruption (see Fig. 1) and the fluid instabilities associated with the falling material were described recently by \citet{2012A&A...540L..10I}.  \citet{2013ApJ...764..165W} estimated the mass of the rapidly falling prominence material using high cadence EUV images from SDO/AIA, while \citet{Reale20062013} used a combination of observations and simulations to investigate solar surface brightening due to the falling material. This event was also shown by \citet{2012ApJ...746...13L} to trigger a globally propagating EUV wave. 

 The focus of the present paper is the interaction between falling material from the dramatic partial filament eruption and the solar atmosphere as the material returns to the surface.  We address the physical mechanisms that are potentially responsible for the observed brightenings, with the objective of determining which mechanism is dominant. In this context we utilize imaging observations from both SDO/AIA and STEREO/EUVI. In doing so we address the following question: are the observed EUV brightenings caused by the prominence material dissipating its kinetic energy in the chromosphere via collisions (compression), with the compressively heated plasma dissipated primarily by heat conduction or are they the result of reconnection occurring between magnetic field lines involved in the impact (reconnection)? The type of reconnection postulated here is between the magnetic field carried by (or frozen into) the falling plasma and the ambient magnetic structure.  Both processes (i.e., compression and reconnection) are plausible but have different emission signatures.

\section{Data}

Prominences are commonly observed above the solar limb in Hα ($\lambda$ = 6563 $\AA$) or He I (10830 $\AA$) and He II (10830 $\AA$) emission.  When seen projected against the solar disk, prominences appear as dark features in chromospheric lines - such as Hα and He I – due to absorption, and also appear above the solar limb and against the disk in extreme ultraviolet coronal lines, such as Fe XII (195 $\AA$). For 195 $\AA$ the observed radiation arises from hydrogen and helium continuum absorption.  The filament eruption on 2011 June 7 that occurred in NOAA AR 11226 (S22W55) was observed in multiple wavelengths by the Atmospheric Imaging Assembly \citep[AIA;][]{2012SoPh..275...17L} onboard the Solar Dynamics Observatory \citep[SDO:][]{2002AGUFMSH21C..01S, 2012SoPh..275....3P}, which takes full-disk images in 10 (E)UV channels at 0.6'' spatial resolution and high temporal cadence of 12 s.  The filament material returning to the surface appeared in absorption in the EUV lines.  It was also observed by the Extreme-Ultraviolet imager \citep[EUVI][]{2004SPIE.5171..111W, 2008SSRv..136...67H} imaging package onboard the Solar-Terrestrial Relations Observatory \citep[STEREO;][]{2008SSRv..136....5K}. EUVI provides observations in four passbands, namely, 171 $\AA$ (Fe IX), 195 $\AA$ (Fe XII), 284 $\AA$ (Fe XV) and 304 $\AA$ (He II). The 195 $\AA$ (Fe XII) data was used for this study.

\section{Analysis and results}

The two mechanisms, “compression” and “reconnection”, are both viable explanations for the observed brightening observed in the SDO/AIA passbands upon impact of in-falling filament material. Clearly, the details of the magnetic topology of the falling material are not well-known; therefore, we use the term “reconnection” generically, indicating plasma heating by magnetic energy release. To determine which mechanism is responsible or dominant, we calculate the kinetic energy of the falling material and compare it with the energy associated with the observed emission.  If the energies are comparable, evidence that the material is dissapating its kinetic energy via collisions, then the more likely mechanism is compression. If the energy release in the emission is larger than the kinetic energy, then reconnection is most likely playing an important role (i.e. the energy release is too large to be explained by compression). Allowing for uncertainties in the energy estimate, we impose a requirement that the energy release must be at least an order of magnitude larger than the kinetic energy to favor “reconnection”. This is reasonable and accounts for large uncertainties in the measurement methods.

Although there is visual evidence that a large amount of prominence material is in-falling and interacting with the solar atmosphere, we restrict our analysis to five of the most accurate measurements and estimates of energy, observed from four distinct pieces of impacting prominence material. The trajectories of the impacting material are illustrated in Figure 1 (see Figure \ref{reg_lightcurves} for precise impact areas).

\subsection{Kinetic energy}

To determine kinetic energy, the velocity and mass of the prominence material impacting the chromosphere must be found.  

\subsubsection{Velocities}

\begin{figure}
\begin{center}
\includegraphics[width=7cm,angle=90, bb = 0 0 576 716]{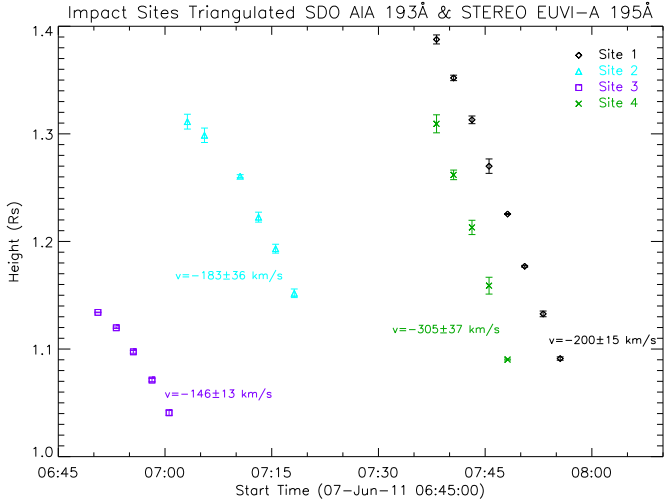}
\caption{Height versus time plots showing the paths of tracked prominence material associated with each impact site (1-4, see Figure \ref{fig1} and Figure \ref{reg_lightcurves}), obtained via trangulation measurements from combining SDO/AIA 193 $\AA$ and STEREO-A/EUVI 195 $\AA$ images. The radial velocity for each piece of material was calculated from the final two height measurements.}
\label{velocity_fig}
\end{center}
\end{figure}

Line-of-sight velocity of the falling prominence material may differ substantially from the true velocity. Also, the presence of magnetic field in the background corona may affect the trajectory of the material, which can deviate from purely ballistic trajectory, and the resulting kinetic energy estimate - proportional to $v^2$ - may differ by an order of magnitude depending on the line-of-sight. To obtain an accurate measure of the true velocities of the falling material (Figure \ref{velocity_fig}), triangulation measurements were performed using observations from the STEREO-A spacecraft, combined with observations from SDO, utilizing the IDL routine in SolarSoft, \verb|ssc_measure| \citep{2009Icar..200..351T}.  This enabled reconstruction of the three-dimensional path of the falling prominence material.  

As Figure \ref{velocity_fig} illustrates, the velocities of material prior to impacting the corona range from approximately 150 km s$^{-1}$ to over 300 km s$^{-1}$.  We use these velocities in conjunction with mass measurements of the pieces of material to obtain an estimate of the kinetic energy associated with the impact.

\subsubsection{Mass}

\citet{2005ApJ...618..524G} developed a technique for deriving prominence mass by observing how much coronal radiation in the Fe XII (195 $\AA$) spectral line is absorbed by prominence material.  In the present work we apply this method, which allows us to consider the effects of both foreground and background radiation in our calculations, to the pieces of prominence falling from the 2011 June 7 eruption to obtain a measure of prominence density. This method also accounts for ``volume blocking'' or the amount of coronal radiation that would be present where the prominence is located.  Volume blocking is described in detail by \citet{2008ApJ...686.1383H} and \citet{2010SSRv..151..243L}. 

The calculation of prominence column mass density (g cm$^{-2}$) along the line of sight requires the determination of the extinction factor (a measure of how much coronal radiation is being absorbed as it travels through a prominence).  If $\sigma$ is the mean absorption cross section for radiation passing through a prominence, the extinction factor for radiation traveling in the direction $\hat{s}$ over a distance $l$ is

\begin{equation}
\alpha = e^{\int^{l}_{0} n \sigma ds },
\label{eqn1}
\end{equation}

where $n$ is the total prominence number density.  If $\sigma$ is uniform throughout the prominence, and if we define the column density by

\begin{equation}
N = \int^{l}_{0} n ds,
\label{eqn2}
\end{equation}

then from Equation \ref{eqn1} and \ref{eqn2} we have

\begin{equation}
N = -\sigma^{-1} \ln \alpha .
\label{eqn3}
\end{equation}

To obtain $\alpha$, we measure intensity in SDO/AIA 193 $\AA$ images in the region of falling prominence material and the region just outside each side of that same prominence material (to interpolate background intensity behind the prominence material).  Due to the nature of the equations, the above mentioned intensity measurements are taken in two adjacent regions characterized by very different background intensities (denoting these regions by the superscripts $L$ and $D$ for light and dark), which occurs at time 07:15 UT as a large piece of material is seen crossing the solar limb \citep[see][for a complete derivation]{2005ApJ...618..524G, 2006ApJ...641..606G} for a complete derivation).

Assuming that the foreground radiation in the light region is simply related to that in the dark region by a proportionality factor that we can specify:

\begin{equation}
I^D_f = \beta I^L_f
\end{equation}

With some manipulation of the equations found in the full derivation, we obtain expressions for the extinction factor and the foreground radiation:

\begin{equation}
\alpha = \frac{I^D_1 - \beta I^L_1}{I^D_0 - \beta I^L_0}
\end{equation}

Applying this technique to the SDO/AIA 193 $\AA$ data at 07:15:45 UT and assigning uncertainties to the various components we find

\begin{equation}
\alpha = 0.028^{+0.0976}_{-0.0206},
\end{equation}

and

\begin{equation}
\ln \alpha = - \left(3.561^{+1.489}_{-1.293}\right).
\label{eqn5}
\end{equation}

Representing the fractional hydrogen and helium abundances (by number) by $f_H$ and $f_{He}$   (where $f_H + f_{He} \approx 1$), and the H and He ionization fractions by $x_H = n_H / (n_H + n_{H^{+}})$ and $x_{He} = n_{He^{+}} / (n_{He} + n_{He^{+}})$, we can write

\begin{equation}
\sigma = f_H (1-x_H) \sigma_H \ + \ f_{He} (1-x_{He}) \sigma_{He} \ + \ f_{He} x_{He} \sigma_{He^{+}}
\label{ion_fracs}
\end{equation}

where $\sigma_H$,$\sigma_{He}$ and $\sigma_{He^{+}}$ are the photoionization cross sections (for 193 $\AA$ radiation) for H, He, and He$^{+}$. The mean prominence mass corresponding to the mean cross section in Equation \ref{ion_fracs} is $m = f_{He}m_{He}  +  f_H m_H = (4f_{He} + f_H)m_H$. 

Taking $\sigma_H$ = 7.69 $\times 10^{-20}$ cm$^{2}$, $\sigma_{He}$ = 1.54 $\times 10^{-18}$ cm$^{2}$ and $\sigma_{He^{+}}$ = 1.01$\times 10^{-18}$ cm$^{2}$ \citep{Kucera2013, 2000asqu.book...95K}, we find from Equation \ref{ion_fracs} that $\sigma = (1.83 \pm 0.4) \times 10^{-19}$ cm$^{2}$ \citep[see Section 5 in][for the range of ionization states considered]{2005ApJ...618..524G}, and from Equation \ref{eqn3} and \ref{eqn5} the column density is

\begin{equation}
N = (1.94^{+1.45}_{-1.01}) \times 10^{19} \ \text{cm}^{-2}.
\label{eqn7}
\end{equation}  

\citet{2013ApJ...764..165W} use multi-wavelength SDO imaging data to apply polychromatic and monochromatic methods in estimating column hydrogen densities of pieces of falling filament material from the same event.  Their estimates are larger $(N_H ~ 10^{20}$ cm$^{-2})$ for a targeted piece of material they describe as exceptional, but are consistent with our results for the thinner threads of material (i.e. lower limit values in the range $N_H \geq 10^{18} - 10^{19}$ cm$^{-2}$), which describe the material analyzed in the current work.  

If the mean mass per particle in the prominence is $m$, then the column mass density $\mu$ is given by $\mu = Nm$, and we can integrate $\mu$ over the prominence area (as seen by the observer) to obtain the prominence mass $M_p$:

\begin{equation}
M_p = m \int \int N da
\end{equation}

In the present work we evaluate $\alpha$ at one point in the prominence, and we must estimate an effective prominence area $A_{eff}$, such that

\begin{equation}
M_p = (4 f_{He} + f_H)m_H \int\int N da \approx (4 f_{He} + f_H)m_H \times N A_{eff}
\label{eqn9}
\end{equation}

Based on the brightening areas of the impacting material (see Table \ref{table1} below), and assuming the blobs of material being measured have the same width and depth dimension, we estimate $A_{eff}$.  We expect this to be an upper limit on the area due to the possible spreading of emission after initial impact. Since $m_H = 1.67\times10^{-24}$ g, it follows from Equation \ref{eqn7} and \ref{eqn9}(7) that the prominence mass range for the tracked piecs of material is

\begin{equation}
M_p = 1.82 \times 10^{12}  -  9.42 \times 10^{13} \ \text{g}.
\label{eqn10}
\end{equation}

Given the range of velocities of the impacting pieces of material in Section 3.1.1 and the estimate of the mass of those pieces in Equation \ref{eqn10}, we obtain the kinetic energy $K.E.$ of the impacting material. We find that $K.E.$ is in the range $7.58 \times 10^{26}  -  5.75 \times 10^{27}$ ergs. The full $K.E.$ estimates and uncertainties are listed in Table \ref{table1}.

\subsection{Observations of plasma impacts and estimation of the radiated energy}

In order to estimate the radiated energy from the observed brightenings at the impact sites, we use the observed AIA fluxes from the brightenings to estimate the differential emission measure (DEM) of the associated plasma volume. This procedure is based on the approach taken by \citet{2013SoPh..283....5A}. The properties of this DEM are used to estimate the total radiated energy per second $dL_{rad}/dt$.

\begin{figure*}
\begin{center}
\includegraphics[width=18.0cm, bb = 0 20 1020 226]{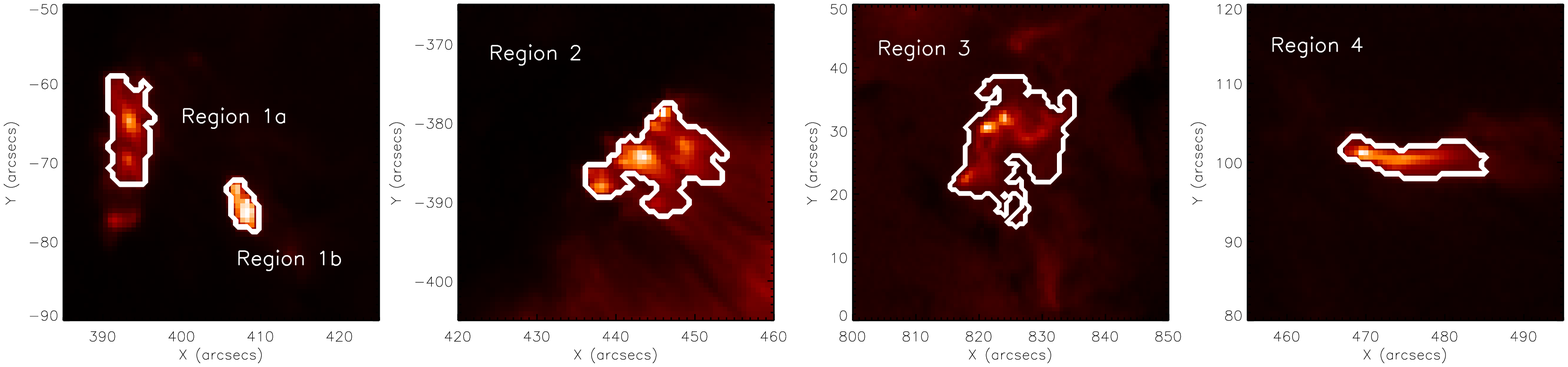}
\includegraphics[width=18.0cm]{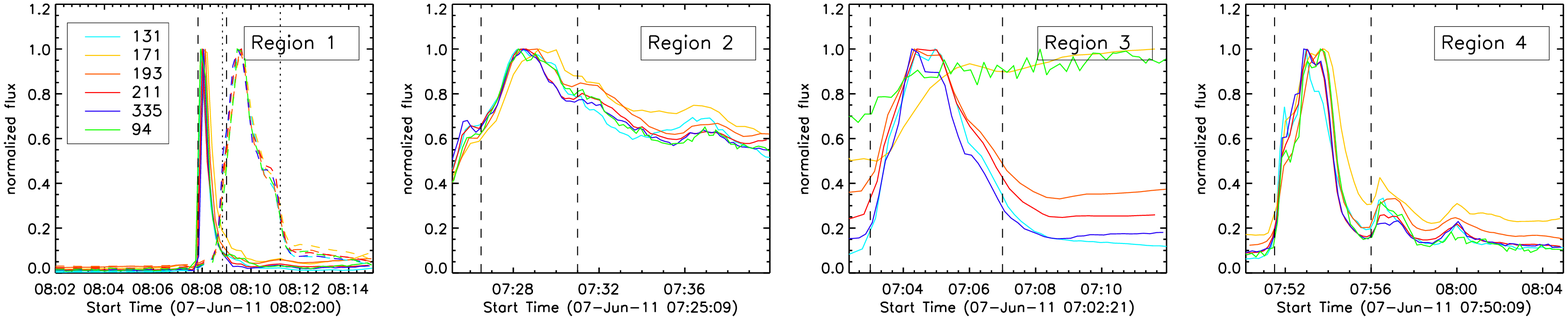}
\caption{Top row: AIA 193$\AA$ images of the chosen brightenings associated with prominence material impact. The contour in each panel shows the defined impact area. Bottom row: AIA flux data from each impact region, as defined in the top panel. All six optically thin EUV wavelengths are shown. }
\label{reg_lightcurves}
\end{center}
\end{figure*}

In this analysis, five of the most clearly observed impacts were selected for study.  Figure 3 shows the location of each of these impacts, two of which (1a and 1b) are associated with the same piece of infalling prominence material. Figure 3 also shows the light curves of each of the labeled impacts in each optically thin AIA wavelength band. %For every impact there is a substantial increase in the observed flux on occurrence of the impact in all AIA channels. 

For each of these impacts, we use the recorded flux in each AIA channel in order to estimate the differential emission measure (DEM) distribution as a function of time for the plasma associated with the impact brightening. As described above, this is achieved via forward modeling, where we choose a distribution of the form,

\begin{equation}
\frac{dEM}{dT} = EM_0 \exp \left(\frac{\log T - \log T_c}{2\sigma^2} \right)
\end{equation}

i.e. a Gaussian emission measure distribution with peak temperature $T_c$ and width $\sigma$, as utilized by e.g. \citet{2011ApJ...732...81A, 2011ApJ...736..102A, 2013SoPh..283....5A}.

\begin{figure}
\begin{center}
\includegraphics[width=4.0cm]{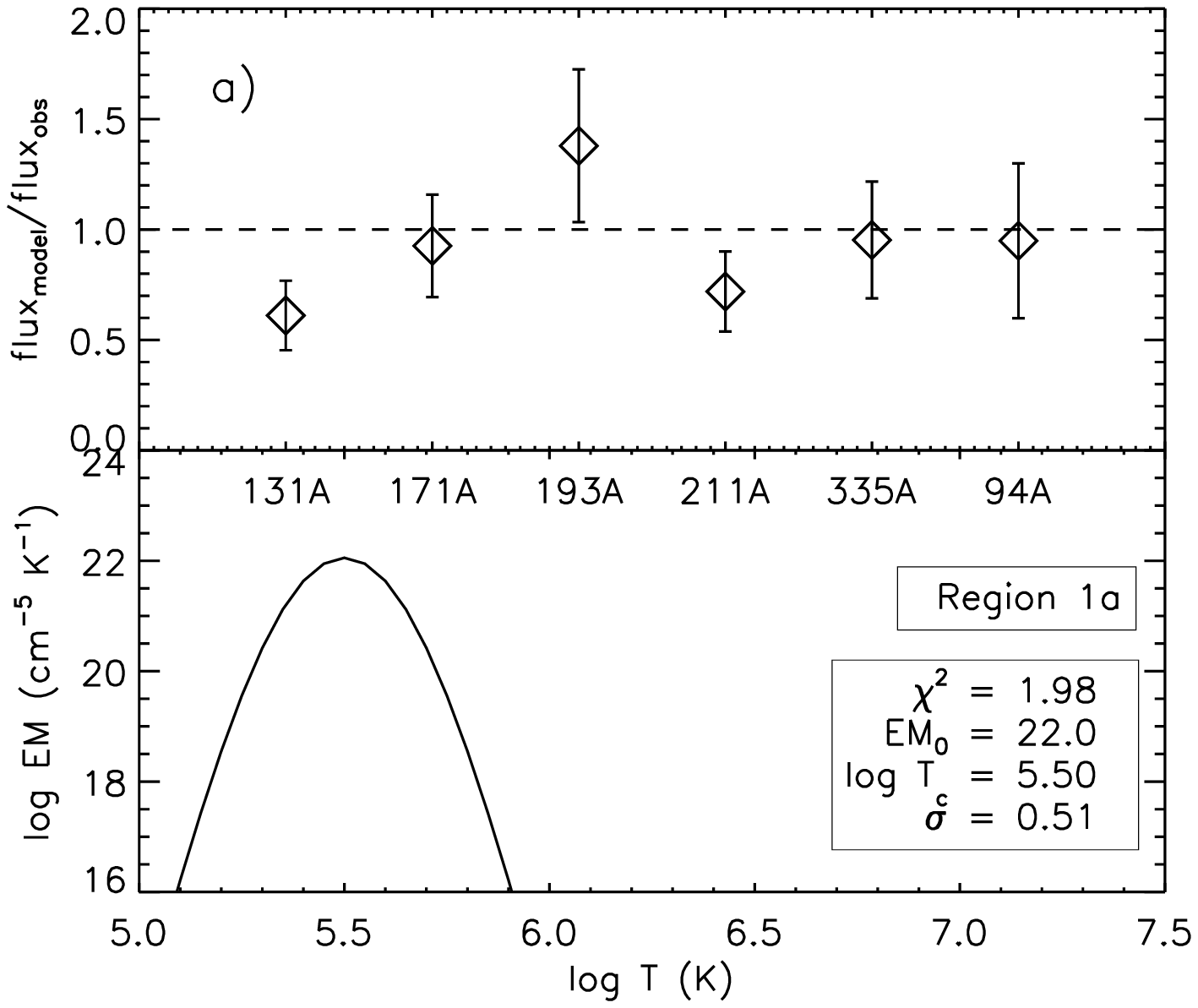}
\includegraphics[width=4.0cm]{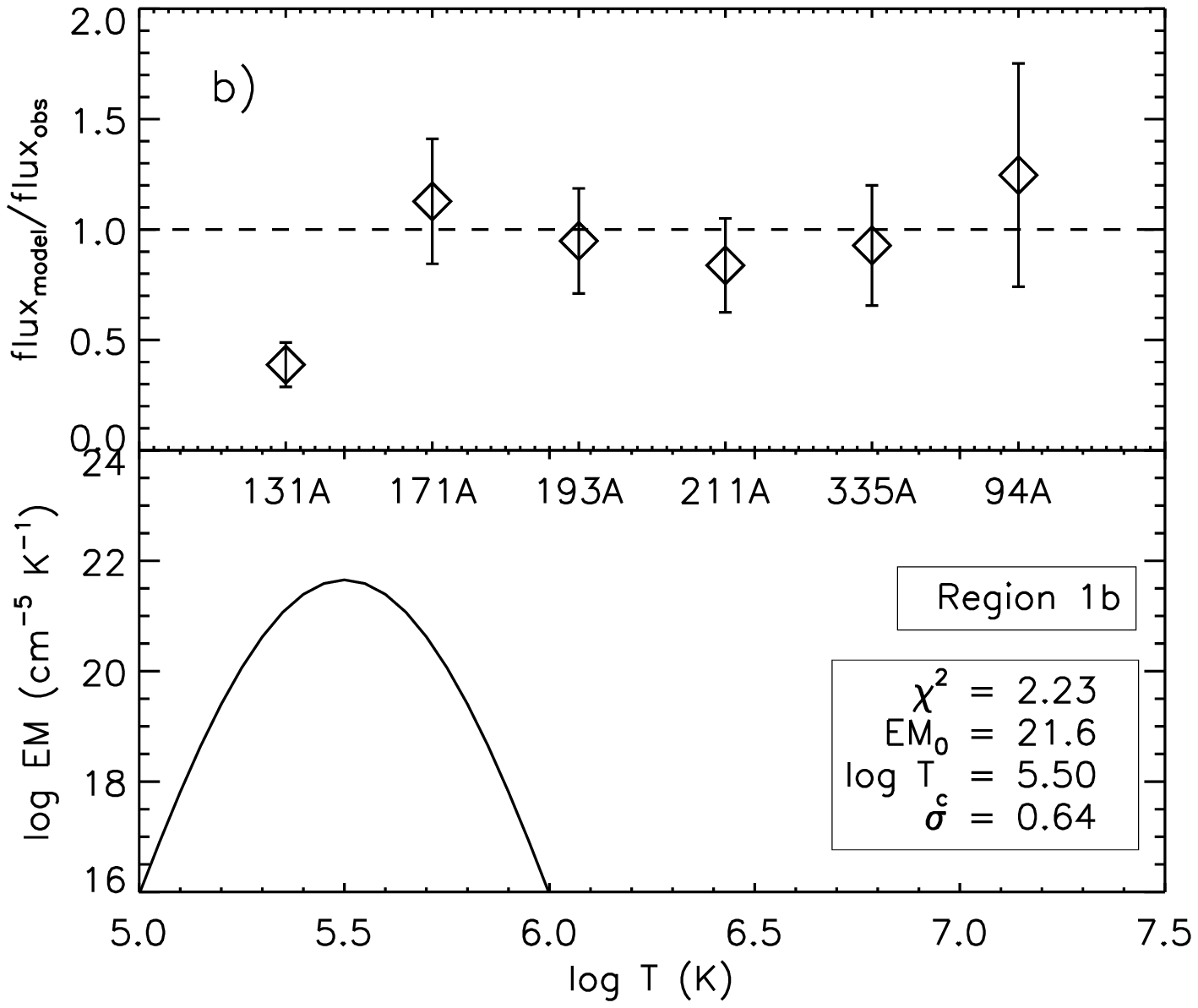}
\includegraphics[width=4.0cm]{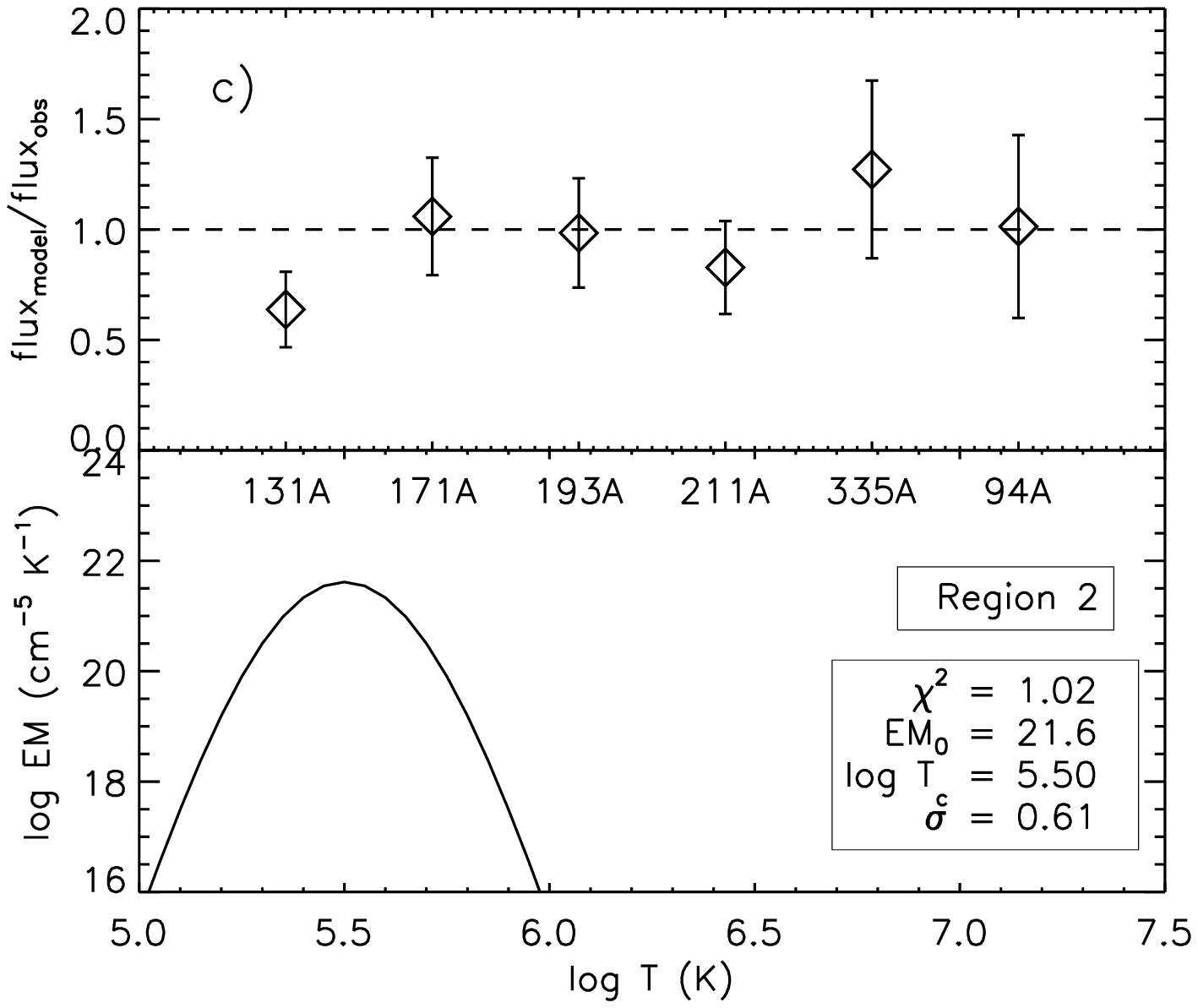}
\includegraphics[width=4.0cm]{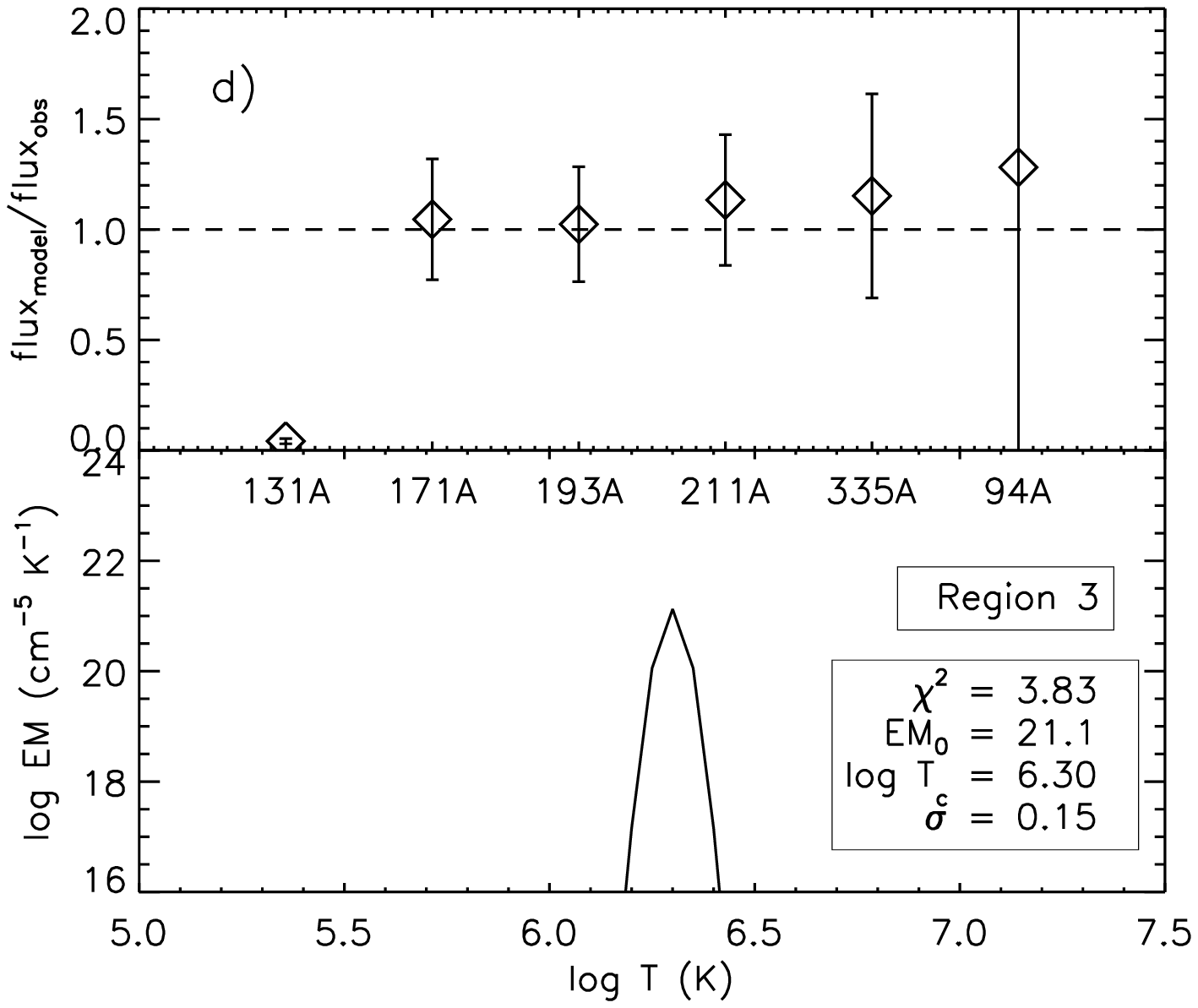}
\includegraphics[width=4.0cm]{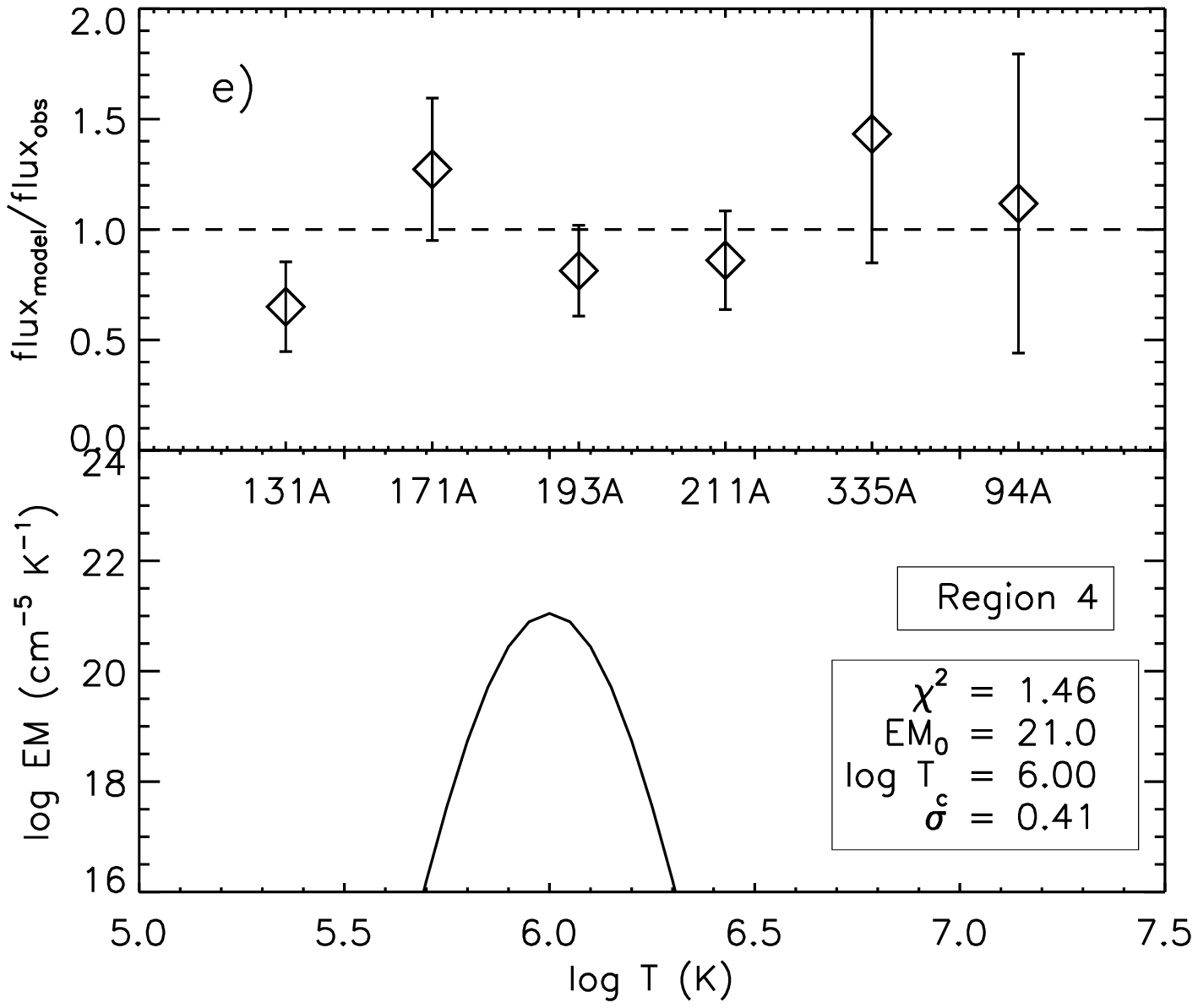}
\caption{Examples of the fitting of differential emission measure distributions to the AIA fluxes from each impact region. The top panels show the ratio of the fitted to the observed flux in each channel; the bottom panels show the best-fit DEM as a function of temperature. }
\label{example_dem}
\end{center}

\end{figure}

The temperature response functions of the AIA channels are the source of significant uncertainty and remain the subject of active study \citep[e.g.][]{2013SoPh..283....5A}, particularly the 94 $\AA$ and 131 $\AA$ channels at low temperatures. In order to account for this, we include a 25\% uncertainty in the measured AIA flux at each wavelength due to instrument response, as suggested by \citet{2012SoPh..275...41B, 2012ApJS..203...26G}. This is combined in quadrature with the statistical uncertainty associated with the AIA flux measurements.

The best fit to the observed flux is achieved at each time interval via a search over the parameter space given by the variables $EM_0$, $T_c$ and $\sigma$ using the $\chi^2$ test. Figure \ref{example_dem} shows an example best-fit for each impact, found in each case near the time of peak EUV emission.

A reasonably constrained DEM allows us to estimate the total radiated energy from the emitting plasma. The radiative loss rate may be written \citep[e.g.][]{2005psci.book.....A}, 

\begin{equation}
\frac{dL_{rad}}{dt} = \int^{T_2}_{T_1} EM(T) \times \Lambda(T) \ dT \ \text{erg s}^{-1}
\end{equation}
\

where $\Lambda(T)$ represents the radiative loss function and $EM(T)$ is the emission measure multiplied by the emitting area $A$, and hence is in units of cm$^{-3}$. For each emitting region $A$ is estimated from the AIA observations of the brightening extent (see Figure \ref{reg_lightcurves}. The radiative loss function $\Lambda(T)$ has been previously estimated by many authors, and varies primarily as a function of temperature \citep{2005SoPh..227..231W}. For this work, we obtain an estimate of $\Lambda(T)$ from the CHIANTI database \citep{1997A&AS..125..149D, 2012ApJ...744...99L}, choosing appropriate coronal abundances.

The last step towards estimating the total radiated energy is to integrate over the time duration of the event, hence,

\begin{equation}
L_{rad} = \int^{t_1}_{t_0} \frac{dL_{rad}(t)}{dt} dt.
\end{equation}

For each impact, the dashed lines in Figure \ref{reg_lightcurves} denote the start and end times of the integration. Following this procedure, we find that the total radiated energy of these events is $10^{25} - 10^{26}$ ergs. The full results are summarized in Table \ref{table1}.

\begin{table*}
\begin{center}
\scalebox{1.0}{
\begin{tabular}{cccccccc}
\tableline
Impact & Start time & End time & Solar location & Mean impact area & Radial velocity of & K. E. & $L_{rad}$ \\
 & (UT) & (UT) & (arcsec.) & (10$^{17}$ cm$^{2}$) & material (km/s) & (10$^{27}$ ergs) & (10$^{25}$ ergs) \\
\tableline
\tableline
1a & 08:07:50 & 08:09:00 & (407,-75) & 0.9 & -200 $\pm$ 15 & 0.76$^{+0.72}_{-0.45}$ & 1.9$^{+1.3}_{-0.7}$ \\
1b & 08:08:50 & 08:11:12 & (393,-65) & 3.2 & -200 $\pm$ 15 & 2.70$^{+2.76}_{-1.59}$ & 8.4$^{+3.7}_{-2.0}$\\
2 & 07:26:30 & 07:31:00 & (445,-385)& 6.6 & -183 $\pm$ 36 & 4.66$^{+7.04}_{-3.22}$ & 24.7$^{+7.7}_{-4.1}$  \\
3 & 07:03:00 & 07:07:00 & (825, 25) & 12.8 & -146 $\pm$ 13 & 5.75$^{+6.15}_{-3.47}$ & 4.4$^{+1.4}_{-0.8}$ \\
4 & 07:51:30 & 07:56:00 & (475,100) & 2.8 & -305 $\pm$ 37 & 5.49$^{+6.61}_{-3.46}$ & 2.8$^{+0.9}_{-0.5}$ \\

\end{tabular}
}
\caption{Summary of impact region properties}
\label{table1}
\end{center}
\end{table*}

To estimate the uncertainty associated with the total radiated energy estimates, a number of monte-carlo simulations were performed on the calculation of $dL_{rad}/dt$. For each simulation run, one thousand best-fit DEMs were generated based on simulated AIA input fluxes. These were sampled from a distribution generated using AIA flux measurements of the impacts and their uncertainties. The resulting uncertainties on the radiated energy estimates are shown in Table \ref{table1}.

\section{Discussion}

The energy release in emission is 1-2 orders of magnitude smaller than the calculated kinetic energy.  Given our requirement that the energy release must be at least an order of magnitude larger than the kinetic energy to favor “reconnection”, our results show a clear indication that the “compression” mechanism dominates in producing the observed brightenings.  Moreover, we note an observational signature supporting compression over reconnection: all wavelengths in the SDO analysis respond at the same time and have similar decay times.  This is not expected in the case of reconnection since the various lines would show different response times as the plasma cools after being heated.  One might argue that secondary pulsations seen in the AIA data are a signature of flaring, but we note that in several cases the impacting prominence material is extended into multiple pieces that impact the same area, causing an expected temporal variation in the deposition of energy.

Although these results strongly support that one mechanism is dominant over the other, both are likely occurring since the falling material undoubtedly carries frozen-in magnetic flux. Examining more examples of falling prominence material would provide a more complete understanding of the relative importance of the two mechanisms. Studying the temporal evolution of the emission and comparing the emitted energy to the impacting kinetic energy can provide insight on the efficiency of energy conversion and on the properties of the plasma, such as density, filling factors, thermal, and magnetic pressures. By comparing the emissivity and the evolution of these events to emission from flares in events not associated with falling material can help understanding the relative importance of the magnetic dissipation and the compression that occurs through the Lorentz force acting on the flaring plasma ultimately improving our understanding of coronal heating. A three-dimensional MHD model of an active region impacted by falling cool and dense prominence material needs to be developed in order to improve the understanding of these events. These important tasks are left for future studies.

\begin{acknowledgements}
LO would like to thank NASA grants NNX10AN10G, and NNX12AB34G. The authors are grateful to the SDO/AIA science team for making available the data used in this study. We would also like to thank Terry Kucera for insightful conversations regarding cross-sections.
\end{acknowledgements}

\bibliographystyle{apj}

\end{document}